\documentclass{ws-ijmpe}

\begin{document}

\markboth{B. K. Sahu et al.}{The $\alpha$- decay chains of the $^{287,288}$115 isotopes....}

\catchline{}{}{}{}{}

\title{The $\alpha$- decay chains of the $^{287,288}$115 isotopes using
relativistic mean field theory\\
}

\author{B. K. Sahu}
\address{Department of Physics, Manikeswari College, Garh Tumulia,
Sundargarh-770024, India.
\footnote{Mr. B. K. Sahu,
Email: brajesh@iopb.res.in}\\
}

\author{M. Bhuyan}
\address{School of Physics, Sambalpur University, Jyotivihar, Burla-768019, 
India.
\footnote{Mr. M. Bhuyan,
Email: brajesh@iopb.res.in}\\
}

\author{S. Mahapatro}
\address{Department of Physics, Spintronic Technology and Advanced
Research, Bhubaneswar-752050, India.
\footnote{Ms. S. Mahapatro, 
Email: narayan@iopb.res.in}\\
}

\author{S. K. Patra}
\address{Institute of Physics, Sachivalaya Marg,
Bhubaneswar-751 005, India.
\footnote{Dr. S. K. Patra,
Email: patra@iopb.res.in}\\
}

\maketitle

\begin{history}
\received{(received date)}
\revised{(revised date)}
\end{history}

\begin{abstract}
We study the binding energy, root-mean-square radius, and quadrupole
deformation parameter for the synthesized superheavy element (SHE)
$Z=115$, within the formalism of relativistic mean field theory (RMF).
The calculation is done for various isotopes of $Z=115$ element,
starting from $A=272$ to $A=292$. A systematic comparison between the
binding energies and experimental data is made.The calculated binding
energies are in good agreement with experimental result. The results
show the prolate deformation for the ground state of these nuclei.
The most stable isotope is found to be $^{282}$115 nucleus ($N=167$)
in the isotopic chain. We have also studied $Q_{\alpha}$ and $T_{\alpha}$
for the ${\alpha}$- decay chains of $^{287,288}$115.
\end{abstract}

\section{Introduction}

Studies aimed at the identification of new superheavy elements which
contribute to the fundamental knowledge of nuclear potentials and the
resulting nuclear structure. The concept of an ``Island of stability''
existing near the next spherical doubly magic nucleus heavier than
$^{208}$Pb arises in every advanced model of nuclear structure
\cite{expt115}. The elements upto Z = 118 have been synthesized till
today with half-lives varying from a few minutes to milliseconds
\cite{expt115,hof00}. But theoretically predicted center of the island
of stability could not be located. More microscopic theoretical
calculations have predicted various regions of stability, namely Z = 120,
N = 172 or 184 \cite{rutz97,gupta97,patra99} and Z = 124 or 126, N = 184
\cite{cwiok99,cwiok099,kruppa00}. There is a need to design the new experiments
to solve the outstanding problem of locating the precise island of stability
for superheavy elements. Measurements on the ${\alpha}$-decays provide
reliable information on nuclear structure such as ground state energies,
half-lives, nuclear spins and parities, shell effects, nuclear deformation
and shape co-existence
\cite{ren87,hori91,hodge03,lovas98,audi03,ginter03,sewer06,lepp07,xu07}.
Therefore as one of the most important decay channels for unstable nuclei,
${\alpha}$-decay is extensively investigated both experimentally and
theoretically.

Both  non-relativistic (e.g. Skyrme Hartree Fock) theory \cite{cha97,stone07}
and relativistic microscopic mean field formalism (RMF) \cite{sero86,gambhir90}
predict probable shell closures at Z = 114 and 120. Microscopic interaction for
the existence of the heaviest element was estimated by Meitner and Frisch
\cite{meitner39}. Myers and Swiatecki \cite{myers66} estimated the fission
barriers for wide range of nuclei and also far into the unknown region
of superheavy elements. The historical review on theoretical predictions and
new experimental possibilities are given by A. Sobiczewski, F . A . Garrev
and B . N . Kalinkin \cite{kalinkin01}.

A considerable increase in nuclear stability was expected for the heaviest
nuclei with N $>$ 170 in the vicinity of the closed spherical shells,
$Z = 114$ ( or possibly 120, 122 or 126) and N = 184 , similar to the
effect of the closed shells on the stability of the doubly magic $^{208}$
Pb (Z = 82, N = 126) \cite{rutz97,gupta97,patra99}. The change of shape from
spherical to deformed (oblate/prolate) configuration in the $\alpha$-decay
process gives us valuable information about the nuclear structure
properties \cite{oga04,oga004,zagre03,zagre04,zagre004}. The fusion-evaporation reaction
of $^{243}$Am + $^{48}$Ca, leads to the formation of $^291$115 nuclei.
According to the predictions \cite{oga04,oga004}, the 3n- and 4n-
evaporation channels results the odd-odd isotope $^{288}$115 ($N=173$)
and  odd-A isotope $^{287}$115 ($N=172$). Here our basic motivation is
to study the $\alpha$-decay properties of these synthesized isotopes.
It is also worth mentioning that the scientists at Dubna re-performed the
same experiment, where the results are yet to be published \cite{yuri10}.

The relativistic mean field (RMF) formalism is presented in section II. The
results of our calculation are in section III. Section IV includes the
$\alpha$-decay modes of $^{288}$115 and $^{287}$115 isotopes. Summary of our
results is given in section V.

\section{The relativistic mean-field (RMF) formalism}

The microscopic self consistent calculation is now a standard tool to
investigate the nuclear structure.  The starting point of the RMF theory is
the basic Lagrangian \cite{wale74} (The Linear Walecka Model) that describes
nucleons as Dirac spinors interacting with the meson fields. However, the
original Lagrangian of Walecka has taken several modifications to take care
of various limitations and the recent successful relativistic Lagrangian
density for a nucleon-meson many body system \cite{sero86,gambhir90}
is expressed as,

\begin{eqnarray}
{\cal L}&=&\overline{\psi_{i}}\{i\gamma^{\mu}
\partial_{\mu}-M\}\psi_{i}
+{\frac12}\partial^{\mu}\sigma\partial_{\mu}\sigma
-{\frac12}m_{\sigma}^{2}\sigma^{2}\nonumber\\
&& -{\frac13}g_{2}\sigma^{3} -{\frac14}g_{3}\sigma^{4}
-g_{s}\overline{\psi_{i}}\psi_{i}\sigma-{\frac14}\Omega^{\mu\nu}
\Omega_{\mu\nu}\nonumber\\
&&+{\frac12}m_{w}^{2}V^{\mu}V_{\mu}
+{\frac14}c_{3}(V_{\mu}V^{\mu})^{2} -g_{w}\overline\psi_{i}
\gamma^{\mu}\psi_{i}
V_{\mu}\nonumber\\
&&-{\frac14}\vec{B}^{\mu\nu}.\vec{B}_{\mu\nu}+{\frac12}m_{\rho}^{2}{\vec
R^{\mu}} .{\vec{R}_{\mu}}
-g_{\rho}\overline\psi_{i}\gamma^{\mu}\vec{\tau}\psi_{i}.\vec
{R^{\mu}}\nonumber\\
&&-{\frac14}F^{\mu\nu}F_{\mu\nu}-e\overline\psi_{i}
\gamma^{\mu}\frac{\left(1-\tau_{3i}\right)}{2}\psi_{i}A_{\mu} .
\end{eqnarray}
Where $m$ is the bare nucleon mass and ${\psi}$ is its Dirac spinor. Nucleons
interact with the ${\sigma}$, ${\omega}$, and ${\rho}$ mesons. We obtain the
field equations for the nucleon and mesons. The self-consistent iteration
method solved the coupled equations. The c.m. (center of mass) motion energy
correction is estimated by the harmonic oscillator formula
$E_{c.m.} = \frac{3}{4}(41A^{-1/3})$. From the resulting proton and neutron
quadrupole moments, the quadrupole deformation parameter  $\beta_{2}$, as
$Q = Q_n + Q_p = \sqrt{\frac{16\pi}5} (\frac3{4\pi} AR^2\beta_2)$.
The root mean square (rms) matter radius is defined as
$\langle r_m^2 \rangle = {1\over{A}}\int\rho(r_{\perp},z) r^2d\tau$,
where $A$ is the mass number, and $\rho(r_{\perp},z)$ is the deformed
density. The total binding energy and other observables are also obtained
by using the standard relations, given in \cite{gambhir90}. We use the well
known NL3 parameter set \cite{lala97}. This set reproduces the properties
of the stable nuclei and also predicts for those far from the
$\beta$-stability valley. We obtain different potentials, densities,
single-particle energy levels, radii, deformations and the binding energies.
The maximum binding energy corresponds to the ground state for a given
nucleus and other solutions (intrinsic excited state) are also obtained.

\begin{table*}
\caption{The RMF (NL3) results for binding energy BE, two-neutron separation
energy $S_{2n}$, pairing energy $E_{pair}$, the binding energy difference
$\triangle E$ between the ground- and first-excited state solutions, and the
quadrupole deformation parameter $\beta_{2}$, compared with the corresponding
Finite Range Droplet Model (FRDM) results $^{33}$. The energy is in MeV.
}
\renewcommand{\tabcolsep}{0.30cm}
\begin{tabular}{|c|c|c|c|c|c|c|c|c|}
\hline
&\multicolumn{5}{c|}{RMF (NL3) Result}&\multicolumn{3}{c|}{FRDM Result}\\
\hline
Nucleus& BE  & $S_{2n}$ & $E_{pair}$ & $\Delta E$ & $\beta_{2}$ &  BE  & $S_{2n}$ & $\beta_{2}$ \\
\hline
272 & 1944.3 & 16.7 & 17.3 & 6.51 & 0.255 & 1932.8 &      & 0.182 \\
274 & 1961.0 & 16.6 & 16.9 & 6.20 & 0.244 & 1950.3 & 17.5 & 0.192 \\
276 & 1977.2 & 16.3 & 16.3 & 5.87 & 0.232 & 1967.4 & 17.1 & 0.202 \\
278 & 1992.8 & 15.6 & 15.8 & 5.30 & 0.218 & 1983.9 & 16.5 & 0.202 \\
280 & 2008.0 & 15.1 & 15.4 & 4.77 & 0.196 & 2000.3 & 16.4 & 0.053 \\
282 & 2022.8 & 14.7 & 14.7 & 4.15 & 0.182 & 2015.8 & 15.5 & 0.053 \\
284 & 2036.7 & 13.9 & 14.3 & 3.18 & 0.173 & 2030.8 & 15.0 & 0.062 \\
286 & 2049.8 & 13.1 & 14.0 & 2.06 & 0.165 & 2045.2 & 14.4 & 0.071 \\
288 & 2062.5 & 12.7 & 13.7 & 1.23 & 0.152 & 2059.1 & 13.8 & -0.087 \\
290 & 2074.5 & 11.9 & 13.6 & 0.15 & 0.103 & 2072.6 & 13.5 & -0.079 \\
292 & 2086.5 & 11.9 & 13.5 & 0.02 & 0.060 & 2085.7 & 13.1 & -0.061 \\
\hline
\end{tabular}
\label{Table 1}
\end{table*}

\section{Results and Discussion}
Here we investigated the bulk properties like the binding energies (BE),
quadrupole deformation parameters $\beta_{2}$, charge radii ( $r_{ch}$),
pairing energies $E_{pair}$ by using the relativistic Lagrangian with the
successful NL3 force parameter. Earlier, it is reported that most of the
recent parameter sets reproduce well the ground state properties, not only
for stable normal nuclei but also for exotic nuclei far away from the
$\beta$-stability \cite{patra99,gambhir90,lala97,aru05,patra01,patra09}.

\begin{figure}[th]
\vspace*{14pt}
\centerline{\psfig{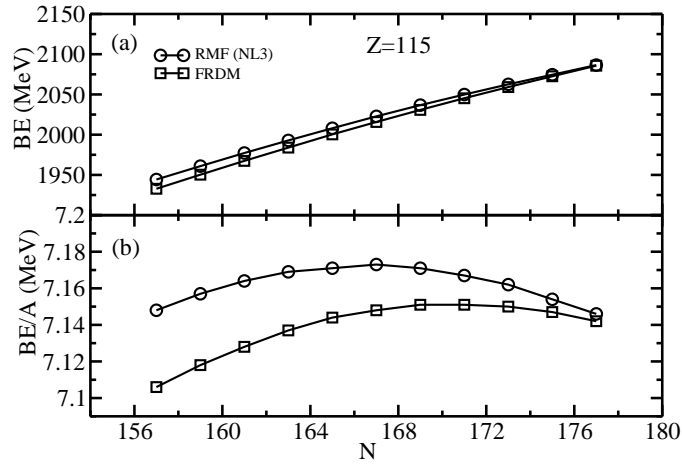}}
\caption{(a)The binding energy BE for the $^{272-292}$115 isotopes, obtained
in RMF (NL3) formalism are compared with the FRDM results $^{33}$.
(b) Same as Fig. 1(a) but for binding energy per particle BE/A.
}
\end{figure}

\subsection{Binding energy and two-neutron separation energy}

The total binding energy ($BE$) for whole isotopic chain for $Z=115$
is plotted in Fig. 1(a) and also listed in Table I. From Fig. 1(a) and
Table I, we notice that the microscopic RMF (NL3) BE over estimated than
that of FRDM at $N = 156-167$, after that the difference in binding energy
decreasing towards the higher mass region (around A=287). And beyond
to this mass number the two curves again showing a similar behaviour.

The binding energy per nucleon ($BE/A$) for the isotopic chain is plotted in
Fig. 1(b). The BE/A value starts reaching a peak value at $A = 282$
for RMF (NL3) and at $A = 286$ for FRDM \cite{moll97,moll097}. It means $^{282}$115
is the most stable isotope from the RMF (NL3) and $^{286}$115 from the FRDM
results \cite{moll97,moll097}. From the above, it is clear that FRDM predicted
$N=171$ closed to predicted closed shell $N = 172$
\cite{rutz97,gupta97,patra99}, which is not appear in case of RMF (NL3).

\begin{figure}[th]
\vspace*{14pt}
\centerline{\psfig{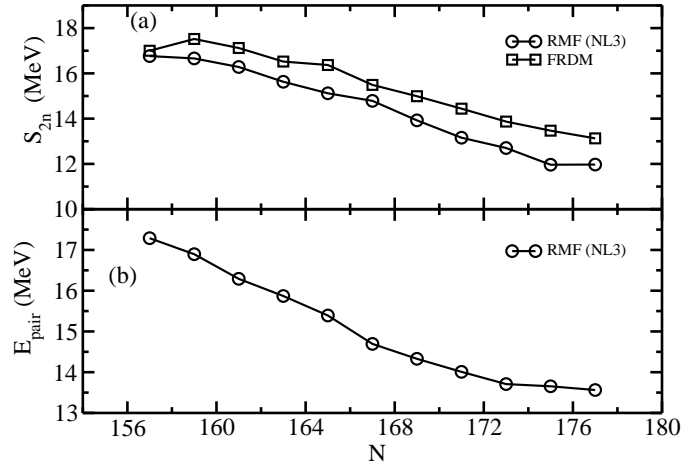}}
\caption{The two-neutron separation energy ${S_{2n}}$ for $^{272-292}$115
nuclei, obtained from RMF (NL3) formalisms, and compared with the FRDM
results $^{33}$, wherever available. (b) The pairing energy $E_{pair}$, for the
relativistic RMF (NL3) formalism.}
\end{figure}

The two neutron separation energy $S_{2n}$ (N, Z) = $BE$ (N, Z) - $BE$
(N-2, Z) is mentioned in Table I. The comparisons of $S_{2n}$ for the RMF
and FRDM models are shown in Fig. 2(a), which shows that the two $S_{2n}$
values coincide remarkably well. $S_{2n}$ values decrease gradually with
increase of the neutron number, except for the noticeable kinks at $A = 282$
(N=167) in RMF and there is no such behaviour in FRDM.

Pairing is important for open shell nuclei whose value, for a given nucleus,
depends only marginally on quadrupole deformation parameter $\beta_2$.
$E_{pair}$ is shown in Fig. 2(b) for the RMF (NL3) calculation, It is
clear from Fig. 2(b) that $E_{pair}$ decreases with increase in mass number
A, i.e, even if the $\beta_{2}$ values for two nuclei are the same, the
$E_{pairs}$ are different from one another. While comparing the results
of paring energy obtained from semi-empirical-mass formula with emperical 
value of the average pairing gap $\Delta \sim 12.A^{-1/2}$, the pairing
energy $E_{pairs}$ from RMF (NL3) calculations overestimated than that 
of the empirical values, saying the failure of extrapolation to SHE 
region of the phenomenological formula.

\subsection{Quadrupole deformation parameter}

The quadrupole deformation parameter $\beta_2$, for both the ground and first
excited states, are also determined within the RMF formalism. In some of the
earlier RMF calculations, it was shown that the quadrupole moment obtained
from these theories reproduce the experimental ground state (g.s). data
pretty well
\cite{patra99,cha97,sero86,gambhir90,lala97,aru05,reinhard95,cha98,brown98}.
The g.s. quadrupole deformation parameter $\beta_2$ is plotted in Fig. 3(a)
for RMF, and compared with the FRDM results \cite{moll97,moll097}. It is clear from
this figure that the FRDM results differ from the RMF (NL3) results for some
mass regions. For example, the prolate structure  has been found for all the
isotopes within RMF. There is a shape change from prolate to oblate at
$A = 286$ (N = 171) to $A = 288$ (N = 173) in FRDM.

\begin{figure}[th]
\vspace*{14pt}
\centerline{\psfig{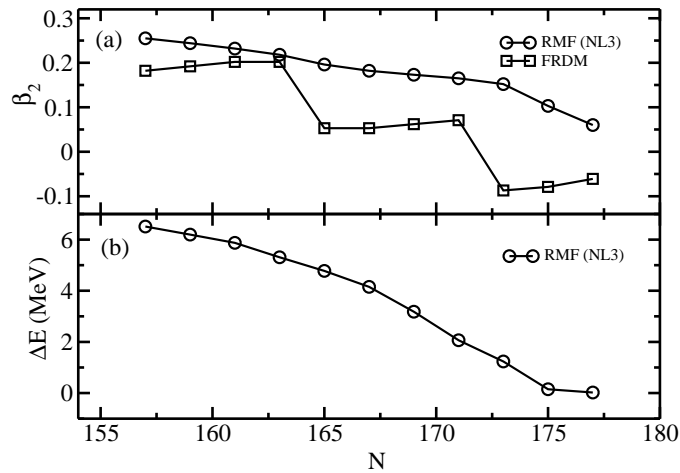}}
\caption{(a) Quadrupole deformation parameter obtained from relativistic
mean field formalism RMF (NL3), compared with the FRDM results $^{33}$,
whereever available. (b) The energy difference between the ground-state
and the first excited state $\triangle E$ compared with the FRDM results
$^{33}$.
}
\end{figure}

\subsection{Shape co-existence}

The binding energy difference $\triangle E$ is the energy difference between
the ground state (g.s.) and the first excited state (e.s.). $\triangle E$ is
plotted in Fig. 3(b). From Fig. 3(b), we notice that $\triangle E$ decreases
with increase in mass number A in the isotopic series. There is a small
difference in binding energy with increase in neutron number. It is an
indication of shape co-existence. For example, in $^{290}$115 the two solutions
for $\beta_{2}$ = $0.103$ and $\beta_{2}$ = $-0.176$ are completely degenerate
with binding energies of $2074.53$ and $2074.38$ MeV. This result shows that the 
g.s.can be changed to the e.s. and vice-versa, by a small change in the input
like the pairing strength etc. in the calculations. Such a phenomenon exists
in many other regions \cite{patra03,patra003,patra0003,patra00003} of the periodic table.

\begin{figure}[th]
\vspace*{14pt}
\centerline{\psfig{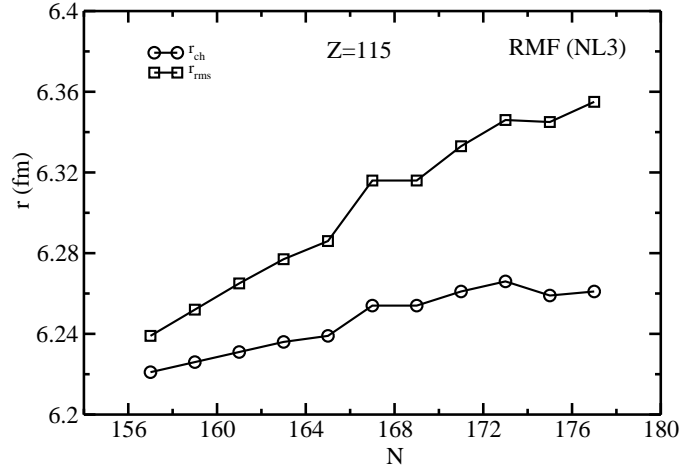}}
\caption{The rms radii $r_m$ of matter distribution and charge radii $r_{ch}$
for $^{272-292}$115 nuclei, using the relativistic mean field formalism RMF(NL3).
}
\end{figure}

\subsection{Nuclear radii}

The root-mean-square matter radius ($r_{m}$) and charge radius ($r_{ch}$)
for the RMF (NL3) formalism are shown in Fig. 4. It clearly shows that the
rms radius increases with increase of the neutron number. Though the
proton number Z = 115 is constant for the isotopic series, the $r_{ch}$
value also increases with neutron number. Both the radii jump to a lower
value at $A = 282$ ( with N = 167 ).

A detailed inspection of Fig. 4 shows that, in the RMF calculations,
both the radii show the monotonic increase of radii till $A = 293$, with
a jump to a lower value at $A = 290$ (with N = 175). There is no data or
other calculation available for comparisons.

\begin{table}
\caption{The $Q_{\alpha}$ energy and half-life $T_{1/2}^{\alpha}$ for
$\alpha$-decay series of $^{287}$115 nucleus, calculated on the RMF (NL3)
model, and compared with the Finite Range Droplet Model (FRDM) results
$^{33}$, the results of other authors $^{40,41}$, and the
experimental data $^{24}$, wherever available. The experimental
$Q_{\alpha}$ value is calculated from the given $^{24}$ kinetic energy
of $\alpha$-particle. The energy is in Mev.}
\renewcommand{\tabcolsep}{0.55cm}
\begin{tabular}{|c|c|c|c|c|c|c|c|c|}
\hline
A & Z & Ref. & BE & $Q_{\alpha}$ & $T_{1/2}^{\alpha}$ \\
\hline
287 & 115 & Expt. \cite{oga04,oga004}        &         & 10.74      & 32$^{+155}_{-14}$ms \\
    &     & RMF                       & 2056.3  & 11.304     & 0.0158s \\
    &     & FRDM \cite{moll97,moll097}        & 2052.7  & 10.256     & 4.265s \\
    &     & \cite{sili10}             &         & 10.789     & 0.155s \\
    &     & \cite{chow08}             &         & 11.21      & 3.55 ms \\
283 & 113 & Expt. \cite{oga04,oga004}        &         & 10.26      & 100$^{+490}_{-45}$ms \\
    &     & RMF                       & 2039.3  & 10.081     & 5.807s \\
    &     & FRDM \cite{moll97,moll097}        & 2034.6  & 9.346      & 426.57 s \\
    &     & \cite{sili10}             &         & 10.313     & 0.676 s \\
    &     & \cite{chow08}             &         & 11.12      & 1.39 ms \\
279 & 111 & Expt. \cite{oga04,oga004}        &         & 10.52      & 170$^{+810}_{-80}$s \\
    &     & RMF                       & 2021.1  & 9.6        & 26.721s \\
    &     & FRDM \cite{moll97,moll097}        & 2015.3  & 10.93      & 4.365s  \\
    &     & \cite{sili10}             &         & 10.57      & 0.034 s \\
    &     & \cite{chow08}             &         & 11.08      & 0.417 ms \\
275 &101  & Expt.\cite{oga04,oga004}       &         & 10.48      &9.7$^{+4.6}_{-4.4}$ms \\
    &     &RMF                      &2002.4   & 9.47       &15.522s \\
    &     &FRDM  \cite{moll97,moll097}      &1998.3   & 10.07      &0.170s \\
    &     &\cite{sili10}            &         & 10.53      &0.010s \\
    &     &\cite{chow08}            &         & 10.34      &6.36ms \\
271 &107  &Expt.\cite{oga04,oga004}        &         & -          & -      \\
    &     &RMF                      &1983.6   & 9.58       &1.47s \\
    &     &FRDM \cite{moll97,moll097}       &1980.1   & 8.66       &575.43s \\
    &     &\cite{sili10}            &         & -          & -    s \\
    &     &\cite{chow08}            &         & 9.07       & 4.73 s \\
\hline
\end{tabular}
\end{table}
\vspace{12 mm}


\begin{figure}[th]
\vspace*{14pt}
\centerline{\psfig{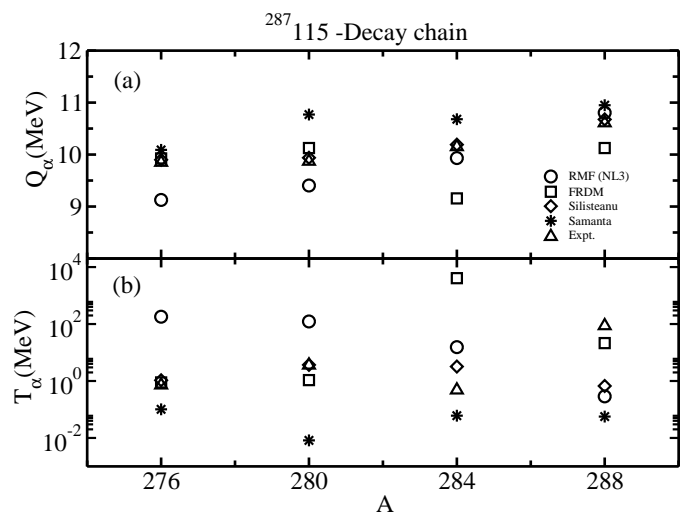}}
\caption{(a) The $Q_\alpha$-energy for $\alpha$-decay series of $^{287}$115
nucleus, using the relativistic mean field formalism RMF (NL3), compared with
the FRDM data $^{33}$, the results of Silisteanu {\it et al.} $^{40}$, Samanta 
{\it et al.}$^{41}$ and the experimental data $^{24}$, wherever available. 
(b) The half-life time $T_{\alpha}$ for $^{287}$115 nucleus using
the RMF(NL3), FRDM, the results of Silisteanu {\it et al.} and Samanta {\it et al.}.
}
\end{figure}

\section{The $Q_{\alpha}$ energy and the decay half-life $T_{1/2}^{\alpha}$}

The $Q_{\alpha}$ energy is obtained from the relation \cite{patra04}:
$Q_{\alpha}(N, Z)$ = $BE (N, Z)$ - $BE (N - 2, Z - 2)$ - $BE (2, 2).$
Here, $BE(N, Z)$ is the binding energy of the parent nucleus with neutron
number N and proton number Z, $BE (2, 2)$ is the binding energy of the
$\alpha$-particle ($^4$He), i.e., 28.296 MeV, and $BE (N- 2, Z - 2)$ is
the binding energy of the daughter nucleus after the emission of an
$\alpha$-particle.

With the $Q_{\alpha}$ energy at hand, we estimate the half-life time
$T_{1/2}^{\alpha}$ by using the phenomenological formula of \cite{parkho05}:
$log_{10}T_{\alpha}^{ph}(Z,N) = aZ [Q_{\alpha}(Z,N) - \overline{E_{i}}]^{-1/2} 
+ bZ + c.$
with $Z$ as the atomic number of the parent nucleus. Where the parameters
${a}$ = 1.5372, ${b}$ = -0.1607, ${c}$ = -36.573 and the parameter
$\overline{E_{i}}$ (average excitation energy of the daughter nucleus) is,

\begin{eqnarray}
\overline{E_i} = &&~~0  ~~~~~~{\rm for}~Z~{\rm even}-N~{\rm even}\nonumber\\
            = && 0.113  ~~{\rm for}~Z~{\rm odd}-N~{\rm even}\nonumber\\
            = && 0.171  ~~{\rm for}~Z~{\rm even}-N~{\rm odd}\nonumber\\
            = && 0.284  ~~{\rm for}~Z~{\rm odd}-N~{\rm odd} .
\label{seqn12}            
\end{eqnarray}

\begin{figure}[th]
\vspace*{14pt}
\centerline{\psfig{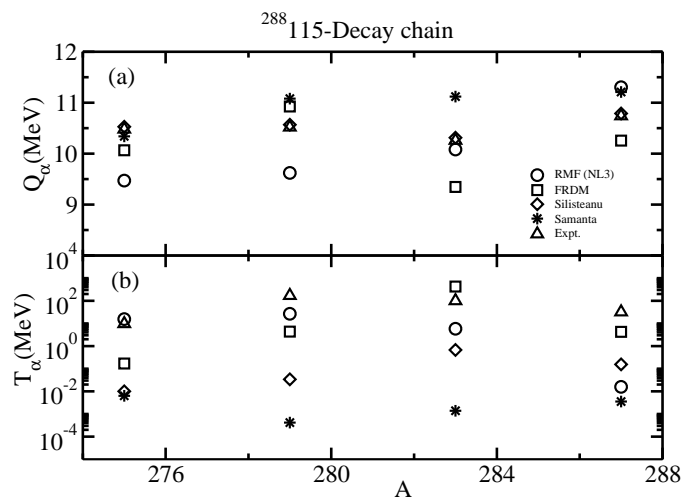}}
\caption{Same as Fig. 5, but for $^{28}$115 nuclear chain.
}
\end{figure}

\subsection{The $\alpha$-decay series of $^{287}$115 nucleus}

We evaluate the BE by using RMF formalism and from these, we estimated
the $Q_{\alpha}$ for whole isotopic chain. We have calculated half-life
time $log_{10}T_{\alpha}$ by using the above formulae. Our predicted results
by using RMF model are compared in Table III with the Finite range droplet
model (FRDM) calculation \cite{moll97,moll097}, the results from Silisteanu {\it at al.}
\cite{sili10}, Samanta {\it et al.} \cite{chow08}, and  experimental data
\cite{oga04,oga004} wherever possible. The comparison of $Q_{\alpha}$ and
$log_{10}T_{\alpha}$(s) are shown in Fig.5(a) and 5(b). From Figure, we notice
that the calculated values of both $Q_{\alpha}$ and $T_{\alpha}$ agree well
with the result of Silisteanu {\it et al.}, Samanta {\it et al.} and  experimental data.

\begin{table}
\caption{Same as Table II, but for $^{288}$115 nuclear chain.}
\renewcommand{\tabcolsep}{0.55cm}
\begin{tabular}{|c|c|c|c|c|c|c|c|c|}
\hline
\hline
A & Z & Ref. & BE & $Q_{\alpha}$ & $T_{1/2}^{\alpha}$ \\
\hline
288 & 115 & Expt. \cite{oga04,oga004}               &        & 10.61      & 87$^{+105}_{-30}$ ms \\
    &     & RMF                              & 2063.0 & 10.81      & 0.288s \\
    &     & FRDM \cite{moll97,moll097}               & 2059.1 & 10.13      & 21.37s \\
    &     & \cite{sili10}                    &        & 10.68      & 0.668s \\
    &     & \cite{chow08}                    &        & 10.95      & 0.056 s \\
284 & 113 & Expt. \cite{oga04,oga004}               &        & 10.15      & 0.48$^{+0.58}_{-0.17}$s \\
    &     & RMF                              & 2045.5 &  9.93      & 15.416s \\
    &     & FRDM \cite{moll97,moll097}               & 2041.0 & 9.16      & 4073.802 s \\
    &     & \cite{sili10}                    &         & 10.19      &    3.206 s \\
    &     & \cite{chow08}                    &         & 10.68      &    0.0605 s \\
280 & 111 & Expt. \cite{oga04,oga004}               &        & 9.87       &  3.6$^{+4.3}_{-1.3}$s \\
    &     & RMF                              & 2027.1 & 9.40       & 123.104 s \\
    &     & FRDM \cite{moll97,moll097}               & 2021.8 & 10.13       &1.0715 s \\
    &     & \cite{sili10}                    &        & 9.94       & 3.68 s \\
    &     & \cite{chow08}                    &        & 10.77      & 0.0082 s \\
276 & 109 & Expt. \cite{oga04,oga004}               &        & 9.85       & 0.72$^{+0.87}_{-0.25}$s\\
    &     & RMF                              & 2008.2 &9.13       & 180.267s \\
    &     & FRDM \cite{moll97,moll097}               & 2003.6 &9.93       & 0.89s \\
    &     & \cite{sili10}                    &        &9.90       & 1.061 s \\
    &     & \cite{chow08}                    &        &10.09       &0.101 s \\
272 & 107 & Expt. \cite{oga04,oga004}               &        & 9.15       &  9.8$^{+11.7}_{-3.5}$s\\
    &     & RMF                              & 1989.0  & 9.36       &  6.74s  \\
    &     & FRDM \cite{moll97,moll097}               & 1985.3  & 8.89       & 229.086s \\
    &     & \cite{sili10}                    &        &  -       & 24.1s \\
    &     & \cite{chow08}                    &        & 9.08     & 16.5 s \\
\hline
\hline
\end{tabular}
\end{table}

\subsection{The $\alpha$-decay series of $^{288}$115 nucleus}

From the BE, which have calculated from RMF formalism, we evaluated $Q_{\alpha}$
and $log _{10}T_{\alpha}$(s) for whole isotopic chain.The predicted results
are compared with FRDM predictions \cite{moll97,moll097}, Silisteanu et.al. \cite{sili10},
Samanta et.al. \cite{chow08}, experimental data \cite{oga04,oga004}, wherever possible.
From Fig. 6(a), 6(b) and Table II, we found that RMF results agree well with
the results of Silisteanu {\it et al.} and Samanta {\it et al.} and the experimental
data.

\section{Summary}
We have calculated the binding energy, rms charge and matter radii, quadrupole
deformation parameter of the isotope of $^{287}$115 and $^{288}$115 and also
investigated two-neutrons separation energy and energy difference between
ground and first excited state, for studying the shape co-existence, pairing
energy, for the isotopic chain of Z = 115. We observed the most stable
isotope is $^{282}$115. The value of $Q_{\alpha}$ and $T_{\alpha}$ are in
good agreement with the available experimental data. We have seen that the
RMF theory provides a resonably good description for whole isotopic chain.

\section{Acknowledgments}

Two of the authors (MB and BKS) thank Institute of Physics for hospitality.
This work is supported in part by the UGC-DAE Consortium for Scientific
Research, Kolkata Center, Kolkata, India\\
(Project No. UGC-DAE CRS/KC/CRS/2009/NP06/1354).

\end{document}